

\magnification=1200
\hsize=15.5truecm
\vsize=21.5truecm   
\hoffset=.4truecm

\baselineskip=13pt plus 1pt minus 1pt
\parskip=0pt
\parindent=20pt

\def\normal{\baselineskip=13pt plus 1pt minus 1pt}

\font\fiverm=cmr5

\font\ninerm=cmr9

\font\nineit=cmti9

\font\ninebf=cmbx9

\def\myref#1{$^{#1}$}

\def\short#1{\hbox{$\kern .1em {#1} \kern .1em$}}

\def\big{\displaystyle \strut }

\def\N{\kappa}

\def\L{ {\cal L} }

\def\E{ {\cal E} }
\def\O{{\rm O}}

\def\ep{\epsilon}
\def\eps{\varepsilon^{\mu\nu\rho}}
\def\d{\partial}

\def\la{\raise.16ex\hbox{$\langle$} \, }
\def\ra{\, \raise.16ex\hbox{$\rangle$} }
\def\st{\, \raise.16ex\hbox{$|$} \, }
\def\go{\rightarrow}

\def\Tr{{\rm Tr} \, }

\def\eff{ {\rm eff} }

\def\Spect{ {\cal S}\, }

\def\gr{ {\rm g.s.} }

\def\psibar{ \psi \kern-.65em\raise.6em\hbox{$-$} }
\def\Lbar{ {\cal L} \kern-.65em\raise.6em\hbox{$-$} }
\def\Dmu{ D^\mu \kern-1.4em\raise.7em\hbox{$\leftarrow$} ~ }
\def\Dnu{ D^\nu \kern-1.4em\raise.7em\hbox{$\leftarrow$} ~ }

\baselineskip=9pt
\line{\ninerm Preprint from \hfil   UMN-TH-1304/94}
\line{\ninerm University of Minnesota \hfil July 26, 1994}

\vglue 1.5cm

\baselineskip=12pt

\centerline{\bf SPONTANEOUS BREAKDOWN OF THE LORENTZ INVARIANCE}
\centerline{\bf AND THE NAMBU-GOLDSTONE THEOREM\footnote{$^\dagger$}{\nineit
To appear in the Proceedings of DPF'94}}

\vskip .3cm

\baselineskip=12pt

\centerline{\ninerm  YUTAKA HOSOTANI}
\centerline{\nineit School of Physics and Astronomy, University of Minnesota,
               Minneapolis, MN 55455}

\vskip .3cm


{\parindent=15pt  \baselineskip=10pt
\ninerm
\midinsert \narrower\narrower
In a class of three-dimensional abelian gauge theory the Lorentz invariance
is spontaneously broken by dynamical generation of a magnetic field.
An originally topologically massive photon becomes gapless, i.e.\
{\nineit p}$_{\fiverm 0}$=0 at  $\vec {\nineit p}$=0.
Indeed, the photon is
the Nambu-Goldstone boson associated with the spontaneous breaking of
the Lorentz invariance.  Although symmetry generated by two Lorentz boost
generators is broken, there seems to appear only one physical Nambu-Goldstone
boson, namely a photon.  We argue that the Ward identities in  the
Nambu-Goldstone theorem are saturated by the photon pole.
\endinsert
}


\vskip .3cm
\normal

As shown in the previous papers,\myref{1,2}  in a class of
three-dimensional gauge theories described by
$$\eqalign{
\L = &- {1\over 4} \, F_{\mu\nu}F^{\mu\nu} - {\N_0 \over 2} \,
\eps A_\mu \d_\nu A_\rho
+ \sum_a  {1\over 2} \, \big[ \,\psibar_a \, , \,
  \big( \gamma^\mu_a (i \d_\mu + q_a A_\mu)
    - m_a \big) \psi_a \, \big] ~, \cr}  \eqno(1) $$
the Lorentz invariance is spontaneously broken by dynamical generation of
a magnetic field $B$.  We have constructed a variational ground state
which has $B \short{\not=}0$ and has a lower energy density than the
perturbative vacuum.  This is the first consistent renormalizable model
in which the Lorentz invariance is spontaneously broken physically.
The variational ground state resembles the ground state in
 the quantum Hall effect.\myref{3,4}

Suppose that a uniform $B \short{\not=}0$ is dynamically generated, which
defines Landau levels.
There is asymmetry in the zero modes: $E_0 \short{=} \ep(\eta q B) \, m$,
where chirality $\eta \short{=}$$ {i\over 2} \,
\Tr \gamma^0\gamma^1\gamma^2 $ $\short{=} \pm 1$.   They exist in either
positive or negative energy states.  We have considered variational ground
states in which these lowest Landau levels  are either empty or
completely filled.   Accordingly a filling factor $\nu_a\short{=} 0$ or $1$ is
assigned.

One can integrate Dirac fields in (1) in a background field
$\bar F_{12} = - B$ with specified filling fractions $\{ \nu_a \}$.
The resulting effective Lagrangian for gauge fields is
$$\eqalign{
\L_\eff[A'] &= - {1\over 2} B^2
 + {1\over 2}\, F'_{0k} \, \ep \,  F'_{0k}
 - {1\over 2}\, F'_{12}\, \chi \, F'_{12}
- {1 \over 2} \, \eps A'_\mu  \, \N \,\d_\nu A'_\rho   + {\rm O}(A'^3) \cr}
   \eqno(2) $$
$A'_\mu$ represents a fluctuation part of $A_\mu$.
$\ep(p)$, $\chi(p)$, and $\N(p)$ ($p_\mu\short{=}i\d_\mu$) summarize
the fermion one-loop  effect, which depend on $B$ and $\{ \nu_a \}$.  In
particular,  $\N(0) - \N_0$ is the induced Chern-Simons term given by
$\N^{\rm ind}_\gr
 = -  \sum_a  \eta_a q_a^2 \, (\nu_a - \hbox{${1\over 2}$})/2 \pi$

Consider a chirally symmetric model consisting of  $N_{\rm f}$ pairs of
$\eta_a\short{=} +$ and $-$ fermions with   $m_a\short{=} 0^+$ and
$q_a\short{>}0$.  Note $\sum_a \eta_a q_a^2\short{=} 0$.  Further
suppose that the condition $\N_0 $$\short{=} \sum_a \eta_a \nu_a q_a^2/2\pi$
is satisfied.  Then the difference in the energy densities of the variational
ground state and perturbative vacuum in RPA is
$$
\Delta \E =  - {\sum \eta_a \nu_a q_a^3 \over 2\pi^3}  \cdot
\tan^{-1} {8 \sum \eta_a \nu_a q_a^2 \over \pi \sum q_a^2}
  \cdot |B|   + \O(|B|^{3/2}).
  \eqno(3)  $$
When the  coefficient of the linear term ($\short{\propto}|B|$) is negative,
the energy density is minimized at $B \short{\not=}0$ so that the  Lorentz
invariance is spontaneously broken.\myref{1,2}  Note $\N(0)=0$.

The Ward identities associated with the spontaneous breakdown of the Lorentz
invariance  are
$$\lim_{p \go 0} p_\rho \, {\rm FT} \,
 \la {\rm T}[{\cal M}^{0j\rho} F_{0k} ] \ra =  - \ep^{jk} \la F_{12} \ra
= \ep^{jk} B ~~.  \eqno(4)  $$
Here ${\cal M}^{\mu\nu\rho}(x)$ is the angular momentum density and FT stands
for a Fourier transform.   Hence nonvanishing $B$ implies
that there must be gapless poles in the correlation function
$\la {\cal M} F \ra$ on the l.h.s..

One might naively expect two Nambu-Goldstone bosons, corresponding to two
broken  Lorentz-boost generators.  We shall show that there seems only one
Nambu-Goldstone boson which couples to both broken generators and saturates
the Ward identities, and that it is the photon.

Plane wave solutions ($\short{\propto} e^{-ipx}$) for photons exist only if
$$\Spect (p) \equiv \ep^2 p_0^2 - \ep \chi {\vec p\,}^2 - \N^2 =0  \eqno(5) $$
which determines the spectrum $p_0(\vec p\, )$.
They satisfy
$${\vec p\,}^2 \, E_j =
 \Big( {i  \N\over \ep}\, p^j  - \ep^{jk} p^k  p^0  \Big) \, B~.
     \eqno(6) $$
In the radiation gauge div$\, \vec A \short{=}0$,
$(A_0 , A_1 , A_2 )=
( \N/\ep , i p^2 , - ip^1 ) ~ e^{-ipx}$.
The important observation is that the photon has only
one physical degree of  freedom.

In terms of symmetric energy-momentum tensors $\Theta^{\mu\nu}$,
${\cal M}^{\mu\nu\rho}\short{=} x^\mu \Theta^{\nu\rho} -x^\nu
\Theta^{\mu\rho}$.
The Ward identities (4) become
$$\lim_{p \go 0} ip_\rho \bigg\{
 {\d\over \d p_0}\, {\rm FT} \, \la {\rm T}[\Theta^{j\rho} F_{0k} ] \ra
- {\d\over \d p_j}\, {\rm FT} \, \la {\rm T}[\Theta^{0\rho} F_{0k} ] \ra
\bigg\}
= \ep^{jk} B ~~.  \eqno(7)  $$
To see the relevance of the photon pole in the identities (7), we recall
$$\Theta^{\mu\nu} = \Theta^{\mu\nu}_{\rm G} + \Theta^{\mu\nu}_{\rm matter}
{}~~~,~~~  \Theta^{\mu\nu}_{\rm G} = - F^{\mu\lambda} F^\nu_{~\lambda}
+ \hbox{${1\over 4}$} g^{\mu\nu}  F_{\rho\sigma} F^{\rho\sigma} ~~. \eqno(8) $$
Due to $\la F_{12} \ra \short{=} - B$,
$\Theta^{\mu\nu}_{\rm G}$ contains terms linear in $B$ and $F'_{\mu\nu}$.  They
are
$$
\Theta_{\rm G}^{00}, \Theta_{\rm G}^{11}, \Theta_{\rm G}^{22}
\sim   B \, B' ~~,~~
\Theta_{\rm G}^{0j} \sim \ep^{jk} B \, E_k'  ~~,~~
\Theta_{\rm G}^{12} \sim   0 ~~.  $$
Therefore, on the l.h.s.\ of Eq.\ (7) we have, for $(jk)\short{=}(12)$,
$$\eqalign{
\lim_{p \go 0} B ~\bigg\{
&-i \Big( p_0 {\d\over \d p_0} - p_1 {\d\over \d p_1} \Big) \,
    \la {\rm T} [E'_2E'_2] \ra  \cr
&-i \Big( p_0 {\d\over \d p_1} - p_1 {\d\over \d p_0} \Big) \,
    \la {\rm T} [B' E'_2] \ra
-i  p_2 {\d\over \d p_1} \,  \la {\rm T} [E'_1E'_2] \ra  \bigg\} \cr}
   \eqno(9) $$
and a similar one for $(jk)\short{=}(21)$.
Note that $p_j$ here denotes $({\vec p}\,)_j = p^j$.
Further
$$\eqalign{
\la {\rm T} [E'_j E'_k ]\ra &= {i\over \Spect (p)} \,
\bigg\{ \delta_{jk} \Big( \chi\, {\vec p \,}^2 + {\N^2\over \ep} \Big)
 - \chi\, p_j p_k + i \ep^{jk}  \N\, p_0 \bigg\}  \cr
\la {\rm T} [B' E'_j ]\ra &= {i\over \Spect (p)} \,
\big( - \ep^{jk} \ep\, p_0 p_k - i \N\, p_j  \big)  \cr}  \eqno(10)  $$

To see the essential part of the mechanism, let us take an approximation
in which we set $\ep(p)=\chi(p)=1$.  In other words, we keep only the
quantum correction to $\N(p)$, which controls the gapless or gapful nature
of photons.
The evaluation of (9) is straightforward, and is  simplified by
taking $p_0 \go 0$ limit first.  The  l.h.s.\ of (7) becomes
$$
 \lim_{\vec p \go 0} ~{{\vec p\,}^2 \over {\vec p\,}^2 + \N (p)^2_{p_0=0} }
 \cdot  \ep^{jk} B  = \ep^{jk}  B
   \eqno(11) $$
since $\N(p)_{p_0=0} \short{=} {\rm O} ({\vec p\,}^2)$.   In this approximation
the photon pole in the gauge field part saturates the Ward identities.
In a general case $\ep(p), \chi(p) \short{\not=}1$, the residue at the pole
acquires corrections.  It is expected that the matter field part of
$\Theta^{\mu\nu}$ gives an additional contribution, again through the photon
pole,   to make the Ward identity (9) satisfied.  If this is the case, the
photon  is the sole Nambu-Goldstone boson associated with the spontaneous
breakdown of the Lorentz invariance.

The one-degree photon couples to both ${\cal M}^{0j\rho}$ ($j=1,2$).
It gives a non-vanishing contribution to the Ward idientity.  The crucial
point is the condition  $\N(0) \short{=}0$.  It follows not only from
the Euler equations, but also from the Nambu-Goldstone theorem.

We remark that the mechanism described in this article works only if
$\N_0 \short{\not=}0$ and the condition $\N(0)\short{=}0$ is satisfied.
For a general value of $\N_0$ more general variational
ground state needs to be considered.  It is an interesting question at which
values of $\N_0$, with a given fermion content, the Lorentz invariance is
spontaneously broken.   What happens if some of fermions are massive?
With an elaboration of the Nambu-Goldstone theorem, we shall come back
to these questions in near future.

\bigskip

{\baselineskip=10pt \ninerm
This work was supported in part
by the U.S.\ Department of Energy under contract no. DE-AC02-83ER-40105.}

\bigskip

\def\ap#1#2#3{{\nineit Ann.\ Phys.\ (N.Y.)} {\ninebf {#1}}, #3 (19{#2})}

\def\ijmpA#1#2#3{{\nineit Int.\ J.\ Mod.\ Phys.} {\ninebf {A#1}}, #3 (19{#2})}

\def\plB#1#2#3{{\nineit Phys.\ Lett.} {\ninebf {#1}B}, #3 (19{#2})}

\def\prD#1#2#3{{\nineit Phys.\ Rev.} {\ninebf D{#1}}, #3 (19{#2})}

\def\cline{\hfil\noexpand\break  ^^J}

\def\myno#1{\item{#1.}}

\leftline{\ninebf References}

\parindent=13pt
\ninerm

\myno{1}  Y.\ Hosotani, \plB {319} {93} {332}.

\myno{2}  Y.\ Hosotani, preprint, UMN-TH-1238/94.

\myno{3}  Our variational ground state  differs from that of P.\ Cea,  \prD
{32}
{85} {2785}.

\myno{4}  Our picture is  different from those of
J.D.\ Bjorken, \ap {24} {63} {174}, and of
  A.\ Kovner and B.\ Rosenstein, \ijmpA {30} {92} {7419}.



\vfil

\end